\documentclass[aps,prd,twocolumn,groupedaddress,nofootinbib]{revtex4-2}
\usepackage{graphicx}
\usepackage{xcolor}
\usepackage{amsmath}
\usepackage{ulem}
\usepackage{hyperref}
\usepackage{aas_macros}
\usepackage{multirow}
\begin{document}
% Use the \preprint command to place your local institutional report
% number in the upper righthand corner of the title page in preprint mode.
% Multiple \preprint commands are allowed.
% Use the 'preprintnumbers' class option to override journal defaults
% to display numbers if necessary
%\preprint{}

%Title of paper
\title{Constraining axionlike particles with invisible neutrino decay using the IceCube observations of NGC 1068}

% repeat the \author .. \affiliation  etc. as needed
% \email, \thanks, \homepage, \altaffiliation all apply to the current
% author. Explanatory text should go in the []'s, actual e-mail
% address or url should go in the {}'s for \email and \homepage.
% Please use the appropriate macro foreach each type of information

% \affiliation command applies to all authors since the last
% \affiliation command. The \affiliation command should follow the
% other information
% \affiliation can be followed by \email, \homepage, \thanks as well.
\author{Bhanu Prakash Pant}
\email{pant.3@iitj.ac.in}
\affiliation{Department of Physics, Indian Institute of Technology Jodhpur, Karwar 342037, India.}
%\email[]{Your e-mail address}
%\homepage[]{Your web page}
%\thanks{}
%\altaffiliation{}
%Collaboration name if desired (requires use of superscriptaddress
%option in \documentclass). \noaffiliation is required (may also be
%used with the \author command).
%\collaboration can be followed by \email, \homepage, \thanks as well.
%\collaboration{}
%\noaffiliation

%\date{\today}

\begin{abstract}
In the beyond Standard Model (BSM) scenarios, the possibility of neutrinos decaying into a lighter state is one of the prime quests for the new-generation neutrino experiments. The observation of high-energy astrophysical neutrinos by IceCube opens up a new avenue for studying neutrino decay. In this work, we investigate a novel scenario of invisible neutrino decay to axionlike particles (ALPs). These ALPs propagate unattenuated and reconvert into gamma rays in the magnetic field of the Milky Way. This is complementary and independent of the previously done studies where gamma rays produced at the source are used to investigate the ALP hypothesis. We exploit the \textit{Fermi}-LAT and IceCube observations of NGC 1068 to set constraints on the ALP parameters. Being a steady source of neutrinos, it offers a better prospect over transient sources. We obtain 95\% confidence level (CL) upper limits on the photon-ALP coupling constant $g_{a\gamma}\lesssim 1.37 \times 10^{-11}$ GeV$^{-1}$ for ALP masses $m_{a} \leq 2 \times 10^{-9}$ eV. Our results are comparable to previous upper limits obtained using the GeV to sub-PeV gamma-ray observations. Moreover, we estimate the contribution from NGC 1068-like sources to diffuse gamma-ray flux at GeV energies under the ALP scenario.
\end{abstract}

% insert suggested keywords - APS authors don't need to do this
%\keywords{}

%\maketitle must follow title, authors, abstract, and keywords
\maketitle
% body of paper here - Use proper section commands
% References should be done using the \cite, \ref, and \label commands

\section{\label{sec:intro}Introduction}
In late 1970, Peccei and Quinn proposed a new $U(1)$ symmetry with an axion as an associated pseudo Nambu-Goldstone boson to address the absence of \textit{CP} violation in quantum chromodynamics (QCD) \cite{peccei1977constraints, peccei1977cp}. Apart from the QCD axions, the existence of axionlike particles (ALPs) is also independently proposed by various Standard Model (SM) extensions of particle physics \cite{svrcek2006axions, arvanitaki2010string}. ALPs are ultralight pseudoscalar bosons and can be understood as a sort of generalization of the QCD axions. Both axions and ALPs couple weakly to SM particles and are potential candidates to make up a significant content of cold dark matter of the Universe \cite{preskill1983cosmology,abbott1983cosmological,dine1983not,sikivie2010dark}. ALPs possess nonvanishing coupling to photons in the presence of an external electromagnetic field, thus inducing a mixing between them. This leads to the conversion of ALPs into photons and vice versa, namely photon-ALP oscillations, similar to flavur oscillations of massive neutrinos. The ALP mass $m_{a}$ and the coupling constant $g_{a\gamma}$ with photons are treated as independent parameters, whereas they are directly related to each other in the case of QCD axions.

The photon-ALP oscillations served as the basis for many experimental searches that have been performed to detect ALPs. So far, no photons have been detected, and several bounds have been placed on ALP parameters by laboratory experiments \cite{zavattini2006, duffy2006, bregant2008, pugnat2008, ballou2015}. For instance, the CERN Axion Solar Telescope (CAST) \cite{zioutas1999decommissioned} experiment has put a stringent constraint of $g_{a\gamma} <$ 6.6$\times$10$^{-11}$ GeV$^{-1}$ at 95\% CL for $m_a \lesssim 0.02$ eV \cite{cast2017}. In the near future, new-generation experiments like International Axion Observatory (IAXO) \cite{armengaud2019}, Any Light Particle Search (ALPS) II \cite{bahare2013}, STAX \cite{capparelli2016}, and ABRACADABRA \cite{ouellet2019} will provide more stringent bounds on the ALP parameter space. 

Another intriguing possibility that has been proposed to perform ALP searches is to look at astrophysical sources with gamma-ray telescopes \cite{hooper2007detecting, mirizzi2007,angelis2007, deangelis2008, simet2008}.  It is well known that the Universe is opaque to photons with $E > 100$ GeV due to their interaction with low energy background photons of extragalactic background light (EBL) or cosmic microwave background (CMB). Due to this interaction, the gamma rays from high-redshift sources undergo pairproduction, initiating an electromagnetic cascade, which severely constrains their propagation length. 
Under the ALP scenario, the transparency of these photons increases drastically, leading to peculiar modulations in their observed $\gamma$-ray spectra. The detection of these very-high-energy (VHE) fluxes by $\gamma$-ray detectors may provide crucial hints on photon-ALP mixing. Many studies have been performed on several galactic and extragalactic sources \cite{meyer2013first, abramowski2013constraints, ajello2016search, meyer2017fermi, liang2019constraints, calore2020, caputo2021, schiavone2021, li2021limits, dessert2022, fiorillo2022, eckner2022, mastrotaro2022, pant2023, galanti2023, huang2023, diamond2023, pant2024} putting bounds on ALP parameter space thus indicating hints of the emergence of some unconventional physics.

The production mechanism of VHE photons and neutrinos through the hadronic channel, either $p\gamma$ or $pp$, is strongly connected to the production of neutral and charged pions. The angular distribution of neutrinos detected by IceCube points to their extragalactic origin. The associated \textit{in situ} VHE gamma rays may be detected by ground-based detectors. Recently, IceCube Collaboration reported an excess of neutrino events from the direction of NGC 1068 \cite{abbasi2022} (see also Ref. \cite{aartsen2020}), a nearby active type-2 Seyfert galaxy with the jet pointing perpendicular to the line-of-sight. The mean number of signal events was reported to be 79$^{+22}_{-20}$, corresponding to 4.2$\sigma$ level of significance, in the energy range of $\sim$1--15 TeV. In the $\gamma$-ray band, \textit{Fermi} Large Area Telescope (LAT) detected emission in the 0.1--300 GeV range \cite{abdollahi2020} but no emission was observed by Major Atmospheric Gamma Imaging Cherenkov Telescopes (MAGIC) and placed strong upper limits in the TeV scale \cite{acciari2019}. Many works have been performed to explain the lack of TeV emission from this source by considering the optically thick environments such as disk corona  \cite{murase2020,inoue2020, murase2022}, wind-dusty torus \cite{inoue2022}, starburst region \cite{eichmann2022}, magnetized corona \cite{kheirandish2021}, etc., thus disfavoring the transparent source scenario.

In this work, we investigate a novel scenario of invisible neutrino decay to ALP. We assume that a fraction of neutrino flux undergoes decay while propagating, producing ultralight ALPs as decay products before reaching the Earth. These ALPs then travel unattenuated throughout the cosmological distance and enter the Galactic magnetic field (GMF), where they again convert into gamma rays and may appear above the detection threshold in the energy range explored by ground-based gamma-ray telescopes. We exploit the \textit{Fermi}-LAT and IceCube observations of NGC 1068 to put 95\% CL constraints on ALP parameter space. This complements previous studies where only the photon-ALP oscillations of gamma rays produced at the source are considered in various magnetic field environments.

This paper is organized as follows. In Sec. \ref{sec:neutrinodecay}, we describe the invisible neutrino decay over the cosmological distance and the probability of ALP production. In Sec. \ref{sec:nudectime}, we examine the current bounds on the neutrino decay lifetime and the bound used in this work. In Sec. \ref{sec:galposc}, we briefly describe the photon-ALP oscillations and conversion in the magnetic field of the Milky Way. In Sec. \ref{sec:nusrc}, we describe the model considered for initial neutrino flux at the source. In Sec. \ref{sec:results}, we place upper limits on ALP parameters. We also discuss the uncertainty introduced to the upper limits due to the choice of initial neutrino flux, GMF models, neutrino decay lifetimes, and $\gamma$-rays to ALPs conversion. Finally, we estimate the diffuse gamma-ray flux at GeV energies from NGC 1068-like sources under the ALP scenario.

\section{\label{sec:neutrinodecay} Invisible neutrino decay}
In the SM of particle physics, the assumption that neutrinos are massless is a direct consequence of the assumption that only left-handed neutrinos exist. Over several decades, many experiments were performed that point to their nonzero mass. The most direct experimental evidence comes from the observation of neutrino oscillations (see Ref. \cite{pdg2022} for a review), which strongly indicate the existence of physics beyond the Standard Model. It is now well known that a neutrino state of specific flavor is a coherent superposition of three mass eigenstates. 

Apart from neutrino oscillation, there are several other BSM scenarios where nonstandard neutrino interactions (NSI) are probed. One such typical scenario is neutrino decay. The nonzero neutrino mass leads to the possibility of heavier neutrinos to decay into lighter ones, which can be either an active (visible decay) or a sterile neutrino (invisible decay) state, along with the emission of a new BSM boson. In 1972, neutrino decay was proposed as a solution to solar neutrino problem \cite{bahcall1972} although it was shown later that it contributes at a subleading order \cite{acker1994}. In invisible decay, and for Majorana neutrinos, $\nu_i \rightarrow \nu_j + a$, where $\nu_i$ and $\nu_j$ are the heavier and lighter neutrino mass eigenstates with masses $m_{i}$ and $m_{j}$, respectively, and $a$ is the pseudoscalar boson of mass $m_{a}$. Note that here we assume $m_{i} > m_{j} + m_{a}$. In this work, we focus on invisible neutrino decay where the pseudoscalar boson is considered an ALP. We mainly follow the methodology developed in Ref. \cite{huang2023}, the relevant expressions of which are recollected below.

The interaction Lagrangian between neutrinos and ALP is given by
 \begin{equation}
     \mathcal{L}_{int} = \sum_{i,j}\text{i}f_{ij}\overline{\nu_{i}}\gamma_{5}\nu_{j} a + h.c. \, , \label{eq:intlag}
 \end{equation}
 where $f_{ij}$ are the couplings in the mass basis.

 The total decay rate is given by \footnote{Here, massive neutrinos are assumed to be Majorana particles. For Dirac neutrinos, the decay rate will be divided by a factor of four, and the coupling should be replaced by $\left| f_{ij} \right|$.}\cite{kim1990, funcke2020}
 \begin{eqnarray}
     \Gamma(\nu_i \rightarrow \nu_j + a) &= \frac{(\text{Re}\,f_{ij})^2 \, m_i}{4\pi} \left[\left(1-\frac{m_j}{m_i}\right)^2-\frac{m^2_a}{m^2_i}\right] \nonumber \\ 
     &\sqrt{\left(1+\frac{m^2_j}{m^2_i}-\frac{m^2_a}{m^2_i}\right)^{2}-\frac{4m^2_j}{m^2_i}}\, . \label{eq:totdec}
 \end{eqnarray}
 It should be noted that in the above equation, the decay rate of channels $\nu_i \rightarrow \nu_j + a$ and $\nu_i \rightarrow \Bar{\nu}_j + a$ are summed up. Also, the decay rate remains the same for antineutrinos.
 
 Assuming normal ordering (NO), i.e., $m_1$ $<$ $m_2$ $<$ $m_3$, with the lightest neutrino $\nu_1$ to be stable, the lifetimes of $\nu_2$ and $\nu_3$ are given by $\tau_2 \equiv \Gamma^{-1}(\nu_2 \rightarrow \nu_1 + a) \nonumber$ and $\tau_3 \equiv \left[\Gamma(\nu_3 \rightarrow \nu_1 + a) + \Gamma(\nu_3 \rightarrow \nu_2 + a)\right]^{-1}$. We take the lightest neutrino $m_1$ to be massless and use the current three-flavour neutrino oscillation data from Ref. \cite{esteban2020fate} to determine $m_2\approx8.61$ meV and $m_3\approx50.1$ meV. We also assume that $m_{2}, m_{3} \gg m_{a}$. 

While propagating over the cosmological distances, neutrinos will decay into ALPs such that their number $N_i(z)$ of mass eigenstate $\nu_i$ changes with redshift $z$ as
\begin{equation}
    \frac{1}{N_i(z)} \frac{dN_i(z)}{dz} = \frac{-1}{\lambda_i(z)} \frac{dl(z)}{dz} \, , \label{eq:nudeceq}
\end{equation}
where $\lambda_i(z) = (1+z)E_{\nu}(\tau_i/m_i)$ is the decay length of neutrino $\nu_i$ at redshift $z$, $E_{\nu}$ is the neutrino energy at $z=0$, and $l(z)$ is the cosmic length traveled by $\nu_i$ from NGC 1068 (at $z_0 = 0.0048$) to redshift $z$. 

On further simplification of Eq. (\ref{eq:nudeceq}), one can obtain the survival probability of neutrinos as the ratio between the number of neutrinos at redshift $z$ and that at NGC 1068,
\begin{equation}
   \frac{N_i(z)}{N_i(z_0)} = \text{exp}\left(\frac{-m_i\,l_{eff}(z)}{\tau_iE_{\nu}}\right) \,, \label{eq:survprob}
\end{equation}
with
\begin{equation}
    l_{eff}(z) = \frac{c}{H_0}\int^{z_0}_{z}\frac{dz}{(1+z)^2} \frac{1}{\sqrt{\Omega_m (1+z)^3+\Omega_{\Lambda}}} \,,
\end{equation}
where $H_0$ is the Hubble constant, $\Omega_m \approx 0.315$, and  $\Omega_{\Lambda} \approx 0.685$. 

Therefore, the ALP flux at the detector ($z=0$) arising from $\nu_i$ decays is given by
\begin{equation}
    \phi_a =  \sum_{\alpha=\mu,e} \mathcal{P}_{\nu_{\alpha}a}(E_{\nu})\,\phi_{\nu_{\alpha}} \, , \label{eq:alpflux}
\end{equation}
where
\begin{equation}
    \mathcal{P}_{\nu_{\alpha}a}(E_{\nu}) = \sum_{i=2,3}\left[1 - \text{exp}\left(\frac{-m_i\,l_{eff}(0)}{\tau_iE_{\nu}}\right)\right]|U_{\alpha i}|^2 \, \label{eq:prodalp}
\end{equation}
is the production probability\footnote{Here, we have neglected the contribution from $\nu_{3}$ decay, i.e., $\nu_3 \rightarrow \nu_2 + a$, and the subsequent $\nu_{2}$ decay.} of ALPs with $l_{eff}(0) = 14$ MPc, $\phi_{\nu_{\alpha}}$ is the neutrino flux of $\nu_{\alpha}$ at the source, and $U_{\alpha i}$ denotes the leptonic flavor mixing matrix \cite{pdg2022}. Here, we consider the standard parametrization of the leptonic mixing matrix and use the best-fit values of neutrino oscillation parameters from Ref. \cite{esteban2020fate}.

\section{\label{sec:nudectime}Current bounds on neutrino decay lifetime}
In Eq (\ref{eq:prodalp}), it can be clearly seen that the ALP production probability depends exponentially on the neutrino decay lifetime. Therefore, it is natural to examine the current bounds in the literature. Various bounds on invisible neutrino decay considering Supernova 1987A \cite{frieman1988}, solar neutrinos \cite{beacom2002, bandyo2003, aharmim2019}, atmospheric neutrinos \cite{gonzalez2008, choubey2018}, and astrophysical neutrinos \cite{denton2018} are proposed (see also \cite{arguelles2023} and the references therein). The bounds on decay lifetimes set by using noncosmological neutrinos are generally weak due to Lorentz suppression of their decay rate since the detectable neutrinos are ultrarelativistic. In this work, we consider the cosmological constraint $\tau_{\nu}\gtrsim 4\times10^{5} \, \text{s} \, (m_{\nu}/50 \,\text{meV})^5$ from Ref. \cite{barenboim2021} for being more stringent over others. Here, $m_{\nu}$ is the mass of the heavier active neutrino. It should be noted that the above constraint applies only to the invisible decay of relativistic neutrinos and needs to be revised for the nonrelativistic case \cite{barenboim2021}.

As mentioned in the previous section, we consider normal mass ordering with $m_1 = 0$, $m_2\approx8.61$ meV, and $m_3\approx50.1$ meV along with the condition $m_{2}, m_{3} \gg m_{a}$. Using these and Eq. (\ref{eq:totdec}), we can estimate the lifetimes as \cite{huang2023}
\begin{subequations}
\begin{eqnarray}
    \tau_{2}&\approx&4\times10^{5} \, \text{s} \, \left(\frac{\text{Re}\, y_{21}}{1.25 \times 10^{-7}}\right)^{-2} \left(\frac{m_2}{50 \,\text{meV}}\right)^5 \, , \label{eq:nu2life} \\
    \tau_{3}&\approx&4\times10^{5} \, \text{s} \, \left(\frac{\text{Re}\, y_{31}}{6.39 \times 10^{-10}}\right)^{-2} \left(\frac{m_3}{50 \,\text{meV}}\right)^5 \, . \label{eq:nu3life}
\end{eqnarray}
\end{subequations}

It is clear from the above equations that to satisfy cosmological constraint, the couplings should be Re $y_{21} \leq 1.25 \times 10^{-7}$ and  Re $y_{31} \leq 6.39 \times 10^{-10}$. We choose Re $y_{21} \sim 10^{-7}$ and  Re $y_{31} \sim 10^{-10}$ which gives $\tau_{2}/m_{2} = 10 ^{4}$ s eV$^{-1}$ and $\tau_{3}/m_{3} = 10 ^{7}$ s eV$^{-1}$ as benchmark values to compute the ALP production probability. It should be noted that much lower coupling values result in an increase in the decay lifetime thus reducing the probability of ALP production. Also, note that $\nu_{2}$ mainly contributes to ALP production since its lifetime is much shorter than that of $\nu_{3}$.

\section{\label{sec:galposc}Photon-ALP Oscillations and conversion in the Milky Way}
The minimal coupling $g_{a\gamma}$ between photons and ALPs $a$ in the presence of an external magnetic field $\textbf{B}$ can be described via the Lagrangian,
\begin{equation}
    \mathcal{L}_{int} = \frac{-1}{4}g_{a\gamma} \,a\,F_{\mu \nu} \tilde{F}^{\mu \nu}=g_{a\gamma}\,a\,\textbf{E} \cdot \textbf{B},
\end{equation}
where $F_{\mu \nu}$ is the electromagnetic field tensor, $\tilde{F}^{\mu \nu}$ is the dual tensor, and \textbf{E} is the electric field of the propagating photon beam. 

Consider a monoenergetic, polarized photon beam of energy $E$ propagating along the $\hat{\textbf{z}}$ direction in a homogeneous external magnetic field \textbf{B} along the $\hat{\textbf{y}}$ axis. In the short wavelength approximation, i.e., $E \gg m_{a}$, the equation of motion can be written as \cite{raffelt1988mixing}
\begin{equation}
\left(i\frac{d}{dz}+E+\mathcal{M}_{0}\right) \psi(z) = 0 \, ,
\end{equation}
with $\psi(z) = \left(A_{x}(z), A_{y}(z), a(z)\right)^T $, where $A_{x}(z)$ and $A_{y}(z)$ denote the photon transverse polarization states along the $x$ and $y$ axis, $a(z)$ is the amplitude associated with ALP field, and $\mathcal{M}_{0}$ represents the photon-ALP mixing matrix.

We assume for simplicity that \textbf{B} is sufficiently weak so that the contribution of the QED vacuum polarization in $\mathcal{M}_{0}$ can be neglected. We also discard the effect of Faraday rotation since we are considering the energy $E$ in the VHE $\gamma$-rays regime. The mixing matrix then takes the simplified form,
\begin{equation}
\mathcal{M}_{0} = 
\begin{pmatrix}
\Delta^{xx} & 0 & 0\\
0 & \Delta^{yy} & \Delta^{y}_{a\gamma} \\
0 & \Delta^{y}_{a\gamma} & \Delta^{zz}_{a}
\end{pmatrix}
\, ,
\end{equation}
with $\Delta^{xx} = \Delta^{yy} = -\omega^{2}_{pl}/2E$, $\Delta^{zz}_{a} = -m^{2}_{a}/2E$, and $\Delta^{y}_{a\gamma} = g_{a\gamma\gamma}B_{y}/2$. Here, $\omega^{2}_{pl}$ is the plasma frequency resulting from the effective photon mass due to the charge screening effect as the beam propagates through the cold plasma.

The propagation length $s$ of the photon-ALP beam can be split into $N$ subregions. The whole transport matrix can be described by $T(s) = T(s_{N}) \times T(s_{N-1}) \times ... \times T(s_{1})$, assuming a constant magnetic field in each subregion. The final survival probability of the VHE $\gamma$-rays on the Earth is $P_{\gamma\gamma} = \text{Tr}\left[(\rho_{11}+\rho_{22})T(s)\rho(0)T^{\dagger}(s)\right]$, where $\rho(0) = \frac{1}{2} \text{diag}(1, 1, 0)$ is the initial polarization of the beam, $\rho_{11} = \text{diag}(1, 0, 0)$ and $\rho_{22} = \text{diag}(0, 1, 0)$ denotes the polarization along the $x$ and $y$ axis, respectively.

In the \textit{strong-mixing} regime, $E_{cr}\leq E \leq E_{max}$, photon-ALP oscillations probability becomes maximal and independent of energy with $ E_{cr} = 2.5 \,\text{GeV} |m_{neV}^2 - 1.4 \times 10^{-3} n_{cm^{-3}}|g_{11}^{-1} B_{\mu G}^{-1}$ and $E_{max} = 2.12 \times 10^{6} \, \text{GeV} \, g_{11} B_{\mu G}^{-1}$. Here, we adopted the notations $m_{neV} \equiv m_{a}/1$ neV, $g_{11} \equiv g_{a\gamma}/10^{-11}$ GeV$^{-1}$, $n_{cm^{-3}} \equiv n_{e}/1$ cm$^{3}$, and $B_{\mu G} = B / 1 \, \mu G$ .

In this work, we consider the ALP-photon conversion in the Milky Way region. The strength of the GMF is $\sim \mathcal{O}$($\mu$G) and can be modeled with both regular (large-scale) and turbulent (small-scale) components. We utilize the GMF model by Jansson and Farrar \cite{Jansson_2012} with updated model parameters according to the Planck satellite data \cite{adam2016planck}. In their previous characterization \cite{Jansson_2012_old}, 22 parameters are used to describe the large-scale regular field as a superposition of disk and toroidal halo components that are parallel to the galactic plane, and an out-of-plane "X-field" component at the galactic center. It also allows the possibility of a striated field and overall scaling of the density of relativistic electrons. These parameters are determined by fitting $\sim$ 40,000 extragalactic Faraday rotation measures (RMs) and polarised synchrotron emission data. In the updated model, 13 more parameters are introduced including the striated and small-scale random fields. Furthermore, the electron density is rescaled, and the scale height is revised. By considering all 36 parameters together, the updated model gives an excellent accounting of the observational data. 

In this study, we consider only the large-scale regular component since in the energy regime of gamma rays, the coherence length of the turbulent component is insignificant to induce any large-scale photon-ALP oscillation effect. We calculated photon-ALP conversion probability in the GMF using publicly available python package \texttt{gammaALPs}\footnote{\url{https://gammaalps.readthedocs.io/en/latest/index.html}} \cite{Meyer:202199}.

\section{\label{sec:nusrc}Initial neutrino flux at the source}
It is evident from Eq. (\ref{eq:prodalp}) that in order to calculate the ALP production probability, a neutrino flux at the source location is required. There are several models proposed in the literature to explain both the \textit{Fermi}-LAT and IceCube observations of NGC 1068 \cite{inoue2020, kheirandish2021, murase2022, inoue2022}. We consider the model by Ref. \cite{eichmann2022} for the benchmark initial neutrino flux at the source. In this model, a two-zone scenario is proposed to explain the broadband multimessenger data by accounting for the nonthermal emission with the inner coronal region and outer starburst ring. The gamma-ray emission above 100 MeV results from the starburst region, whereas the TeV neutrinos originate from the coronal region. In contrast to other models, where an alternative acceleration mechanism is required, it considers cosmic rays (CRs) to be scattered off stochastically by Alfv\'enic turbulence.

\begin{figure*}
    \centering
    \includegraphics[width=\linewidth]{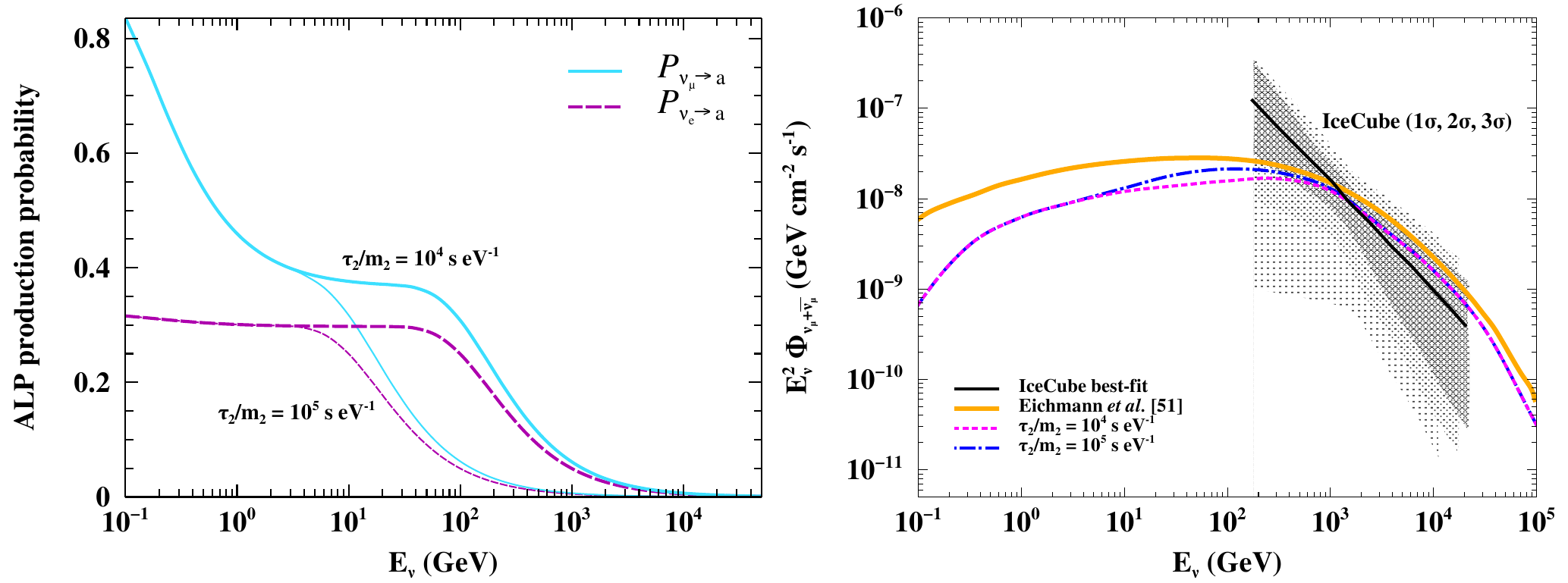}\caption{\label{fig:alpprodprob}Left: ALP production probability from invisible neutrino decay of $\nu_{\mu}$ (solid) and $\nu_{e}$ (dashed) for $\tau_{2}/m_{2}=10^4$ s eV$^{-1}$ (thick) and $\tau_{2}/m_{2}=10^5$ s eV$^{-1}$ (thin) keeping $\tau_{3}/m_{3}=10^7$ s eV$^{-1}$ fixed. Right: survived muon neutrino plus antineutrino flux after the invisible neutrino decay. The initial muon neutrino flux at the source is shown in a yellow solid curve, and the best-fit muon neutrino spectrum of NGC 1068 obtained by IceCube \cite{abbasi2022} is shown in a black solid line along with uncertainties by shaded regions.}
\end{figure*}

 We take the initial neutrino flux $\phi_{\nu}$ from Fig. 5 of Ref. \cite{eichmann2022}, i.e., the total neutrino flux produced in the corona region. Using Eq. (\ref{eq:prodalp}), we calculate the ALP production probability, as shown in the left panel of Fig. \ref{fig:alpprodprob}, from invisible neutrino decay of $\nu_{\mu}$ and $\nu_{e}$ for the benchmark neutrino decay lifetime as described in Sec. \ref{sec:nudectime}. We find that the ALP production probability from $\nu_{\mu}$ decay is $\sim$46\% at 1 GeV, which goes down to below $\sim$5\% at 1 TeV. This implies that around 10 TeV and above, the neutrino decay is negligible, and the neutrino flux nearly matches the benchmark model and the IceCube observation of NGC 1068. In the right panel of Fig. \ref{fig:alpprodprob}, we show the survived muon neutrino plus antineutrino flux after the invisible neutrino decay. For reference, we also show the best-fit muon neutrino spectrum of NGC 1068 by IceCube, along with uncertainties by shaded regions.

\section{\label{sec:results}Results and Discussion}
\subsection{\label{subsec:alpconstraints} Constraints on ALP parameters}
We constrain the ALP parameters by conducting a maximum likelihood analysis on the gamma-ray flux. We construct the likelihood function to be
\begin{equation}
    \mathcal{L} = e^{-\chi^2/2}\, ,
\end{equation}
where $\chi^{2}$ is the chi-squared function.
The $\chi^{2}$ estimator is a function of a single parameter, namely the normalization of ALP flux, and can be written as
\begin{equation}
        \chi^{2}(\theta) = \sum_{i}^{N_{bins}} \left(\frac{\Phi^{\text{conv}}_{i}+\Phi^{\text{$\gamma$ALP}}_{i}(\theta) - \Phi^{\text{obs}}_{i}}{\sigma_{i}}\right)^{2} \, , \label{eq:chi2fit}
\end{equation}
where $\Phi^{\text{conv}}$ is the conventional gamma-ray flux produced due to hadronic interactions and taken from Fig. 5 of Ref. \cite{eichmann2022}, $\Phi^{\text{$\gamma$ALP}}$ is the gamma-ray flux due to conversion of ALPs in the GMF, $\Phi^{\text{obs}}$ are the observed \textit{Fermi}-LAT data points, and $\sigma_{i}$ are the errors associated with the data. The test statistic can be written as
$TS = -2\,\text{ln}\left(\mathcal{L}(\theta)/\mathcal{L}(\hat{\theta})\right)$, which in this particular case reduces to $\Delta \chi^{2} = \chi^{2}(\theta) - \chi^{2}_{min}(\hat{\theta})$, where $\hat{\theta}$ is the best-fit value that minimizes the $\chi^{2}$ function in Eq. (\ref{eq:chi2fit}). We then set upper limits on $g_{a\gamma}$ with 95\% CL by excluding the values for which $\Delta \chi^{2} > 3.84$ for a single degree of freedom \footnote{PDG Review Statistics, Table 40.2} \cite{pdg2022}. 

\begin{figure}
    \centering
    \includegraphics[width=\linewidth]{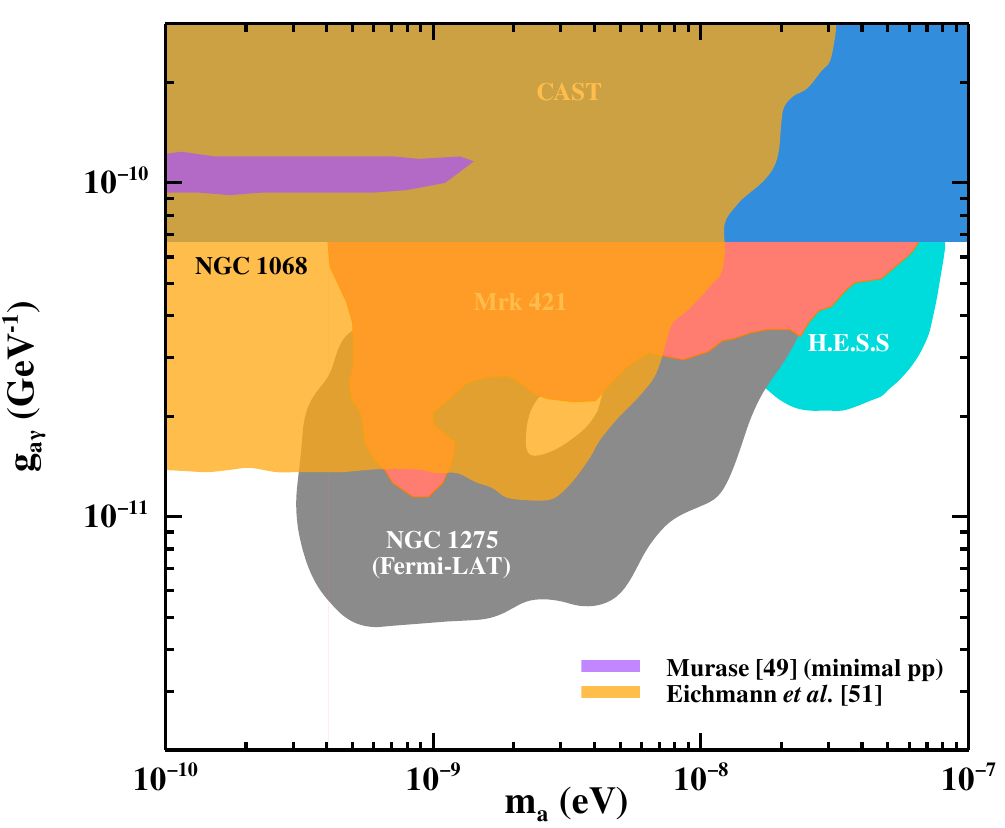}\caption{\label{fig:alplims}Exclusion region at 95\% CL set by NGC 1068 (in yellow) with invisible neutrino decay to ALPs for decay lifetimes $\tau_{2}/m_{2} = 10 ^{4}$ s eV$^{-1}$ and $\tau_{3}/m_{3} = 10 ^{7}$ s eV$^{-1}$. For comparison, we show the constraints set by CAST \cite{cast2017}, NGC 1275 \cite{ajello2016search}, H.E.S.S. \cite{abramowski2013constraints}, and Mrk 421 \cite{li2021limits}. The narrow pink region shows the exclusion region obtained using initial neutrino flux from model II \cite{murase2022} (see text) under the minimal $pp$ scenario. The collection of current upper limits are taken from Ref. \cite{ciaran2020}.}
\end{figure}

We obtain upper limits of $g_{a\gamma} \lesssim 1.37 \times 10^{-11}$ GeV$^{-1}$ for ALP masses $m_{a} \le 2 \times 10^{-9}$ eV as shown in Fig \ref{fig:alplims}. This is a significant improvement over CAST limits; however, it is comparable to existing constraints in this mass range. For comparison, we also show the upper limits set by CAST \cite{cast2017}, NGC 1275 \cite{ajello2016search}, H.E.S.S. \cite{abramowski2013constraints}, and Mrk 421 \cite{li2021limits}. In the left panel of Fig. \ref{fig:gamsed}, we show the ALP-induced gamma-ray spectra resulting from ALP to gamma-ray conversion in the Milky Way corresponding to the upper limits obtained above along with the conventional gamma-ray flux. We find that its strength is too weak to appear as a distinct detectable signature.

\subsection{Uncertainty on upper limits on ALP parameters}
It is implicit from the analysis that the constraint obtained in the previous section may suffer due to uncertainties in various parameters. In this section, we discuss the impact of uncertainty introduced to the upper limits due to the choice of initial neutrino flux, GMF models, neutrino decay lifetimes, and $\gamma$-rays to ALPs conversion.

\subsubsection{Impact of initial neutrino flux}
In this section, we recompute the upper limits on ALP parameters by considering different initial neutrino fluxes as predicted by other models in the literature. These models differ in terms of model parameters, emission region, and acceleration mechanism, thus varying the production mechanism of gamma rays and neutrinos. We choose four different models, as summarized in Table \ref{table:numodels}, for the initial neutrino flux the references to each are given in the last column.

\begin{table*}
\begin{ruledtabular}
\caption{\label{table:numodels}Summary and comparison of neutrino emission mechanism in various models considered for initial neutrino flux.}
\begin{tabular}{c c c c }
 Model no. & Model name & Neutrino emission mechanism & Ref. of $\phi_{\nu}$ \\
 \hline
 I & Kheirandish \textit{et al.} 2021& \multirow{3}{30em}{Both $pp$ and $p\gamma$ interactions; measured intrinsic X-ray luminosities are incorporated in the disk-corona model; CR acceleration via stochastic acceleration and magnetic reconnection are considered. } & Fig. 1 of \cite{kheirandish2021}\\
 & & &\\
 & & & \\
 II & Murase 2022 & \multirow{3}{30em}{Model-independent analysis; (a) $p\gamma$ at $R \lesssim 30 R_S$: accelerated CRs in the corona interact with coronal X-rays and (b) $pp$ at $R \lesssim 100 R_S$: accelerated CRs interact with gas in inflowing material.} &Fig. 3 of \cite{murase2022}\\
 & & & \\
 & & & \\
 III & Inoue \textit{et al.} 2022 & \multirow{3}{30em}{$p\gamma$ interactions in the inner region due to "failed" winds; GeV gamma rays explained via $pp$ interactions of "successful" winds and torus gas in the outer region.} &Fig. 6 of \cite{inoue2022}\\
 & & & \\
 & & & \\
 IV & Blanco \textit{et al.} 2023 & \multirow{3}{30em}{$pp$ interactions; proposed physical model including the acceleration and diffusion of protons; attenuation of VHE gamma rays is described using measured X-ray luminosity.} &Fig. 5 of \cite{blanco2023}\\
 & & & \\
 & & & \\
\end{tabular}
\end{ruledtabular}
\end{table*}

In model I, the measured intrinsic x-ray luminosities are incorporated in the disk-corona model by estimating the relevant parameters to explain the neutrino emission from the magnetized corona. For the CR acceleration mechanism, both stochastic acceleration and magnetic reconnection are investigated.

In model II, a model-independent analysis is done to explain hidden neutrino production with (a) $p\gamma$ scenario at $R \lesssim 30 R_S$, where $R_S$ is the Schwartzschild radius, in which CRs are accelerated in the corona and interact with coronal x-rays and (b) $pp$ scenario $R \lesssim 100 R_S$, in which the accelerated CRs interact with gas in inflowing material.

In model III, protons are assumed to be accelerated via diffusive shock acceleration in the inner regions of the wind near the black hole, thus inducing a nonthermal emission. The $p\gamma$ interactions of "failed" winds are mainly responsible for the neutrino generation. The GeV gamma rays are described by invoking a separate outer region where the "successful" winds interact with the torus gas via $pp$ interactions.

In model IV, a physical model is proposed, including the acceleration and diffusion of high-energy protons, to explain neutrino and gamma-ray emission. The measured x-ray luminosity of NGC 1068 is used to describe the attenuation of VHE gamma rays in the corona region. The emission in the MeV--GeV region arises due to synchrotron and inverse Compton scattering. We choose neutrino flux corresponding to the model parameters $\Gamma_{p}=2.1$, $R = 17 \, R_{s}$, and $B = 6$ kG, where $\Gamma_{p}$ is spectral index of the accelerated protons, $R$ is the radius of the corona, and $B$ is the magnetic field strength in the corona. The model also requires the high-energy electrons to lose energy via synchrotron in the presence of a large magnetic field, $B \gtrsim 6$ kG, to avoid exceeding the observed gamma-ray flux by $Fermi$-LAT at sub-GeV energies.

\begin{figure*}
    \centering
    \includegraphics[width=\textwidth]{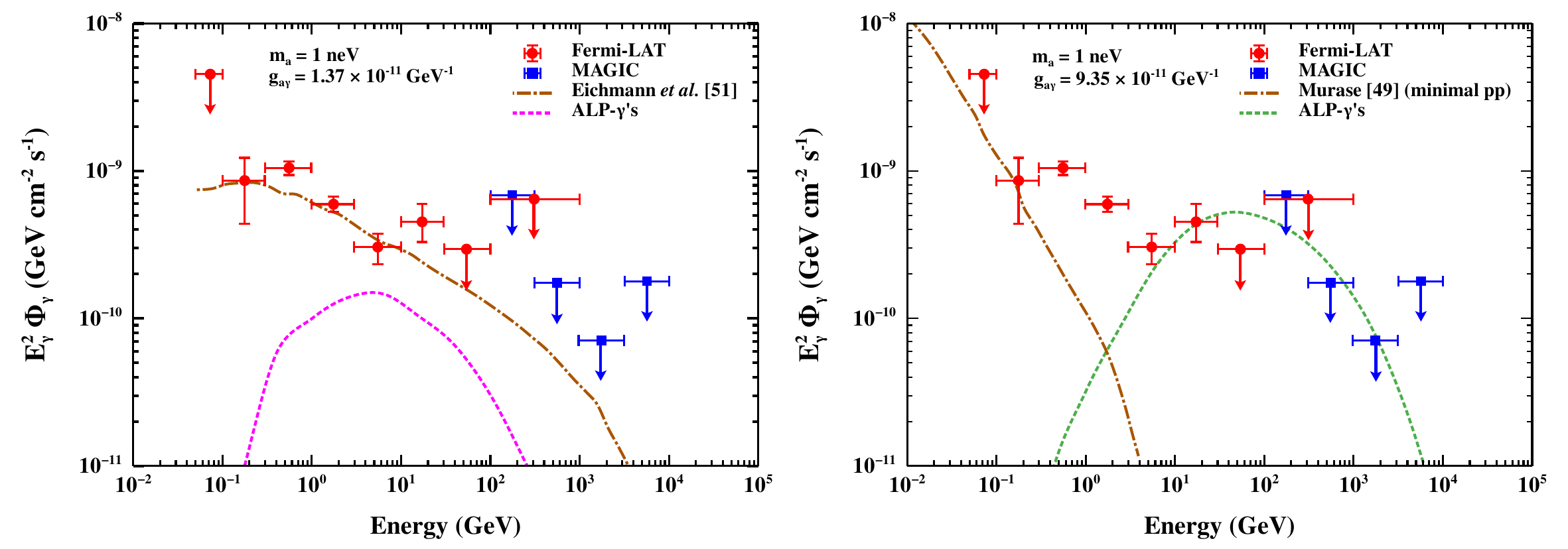}\caption{\label{fig:gamsed}Left: ALP-induced gamma-ray spectra (dotted) corresponding to upper limits of $g_{a\gamma} = 1.37 \times 10^{-11}$ GeV$^{-1}$ for ALP mass $m_{a} = 1$ neV along with the conventional gamma-ray flux (dot-dashed) as estimated in Ref. \cite{eichmann2022}. The \textit{Fermi}-LAT \cite{abdollahi2020} and MAGIC \cite{acciari2019} data points are shown with the red circular and blue square markers, respectively. Right: same as the left panel but for the initial neutrino flux taken from Ref. \cite{murase2022}. The ALP-induced gamma-rays (dotted) are shown for the upper limits obtained in the case of a minimal $pp$ scenario.}
\end{figure*}

We find that no significant constraint can be put on the coupling strength using initial neutrino flux from models I, III, and IV except for the minimal $pp$ case of model II. In this case, a very narrow region of $9.35 \times 10^{-11}\lesssim g_{a\gamma} \lesssim 1.20 \times 10^{-10}$ GeV$^{-1}$ for ALP masses $m_{a} \le 10^{-9}$ eV is excluded, which is shown in Fig. \ref{fig:alplims} as a pink region. This region has already been excluded by the CAST experiment and thus provides no significant improvement. The corresponding ALP-induced gamma-rays are shown in the right panel of Fig. \ref{fig:gamsed}. We infer that the choice of initial neutrino flux plays a significant role in constraining the ALP parameters. In the future, the collection of multimessenger data from a large sample of Seyfert galaxies and the robust analysis of their broadband spectra will lead to a better understanding of the acceleration mechanism, improvements in modeling the emission region, and precisely estimating the model parameters to explain the observed gamma-rays and neutrinos. This will significantly reduce the uncertainty caused by the current incomplete knowledge and allow us to put stringent bounds on ALP parameters. 

\subsubsection{Impact of GMF models}
The probability of conversion of ALPs into photons strongly depends on the magnetic field structure and its strength. In light of the updated measurements in the past few years, the knowledge of the GMF has improved significantly and is more precisely known compared to extragalactic magnetic fields. Nonetheless, we check whether choosing a different GMF model induces a change in the upper limits. We consider an alternate GMF model proposed by Pshirkov \textit{et al.} \cite{pshirkov2011}. In this model, rotation measures of extragalactic sources are used to infer the structure of the GMF.

\begin{figure*}
    \centering
    \includegraphics[width=\textwidth]{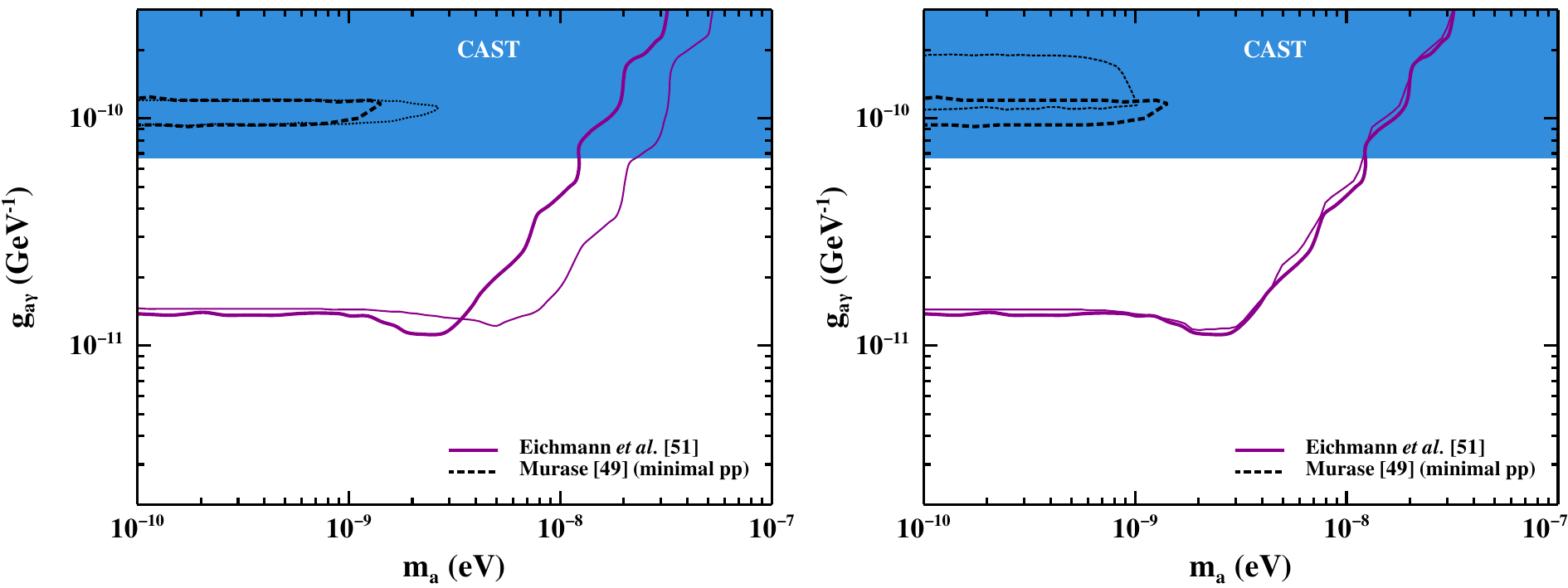}\caption{\label{fig:alplimserr} Left: comparison of constraints on ALP parameters at 95\% CL for the GMF model by Jansson and Farrar (thick) and by Pshirkov \textit{et al.} \cite{pshirkov2011} (thin) for both the benchmark initial neutrino flux and from model II (see text). The neutrino decay lifetime is kept fixed at the benchmark values. Right: same as the left panel but for neutrino decay lifetime $\tau_{2}/m_{2} = 10 ^{5}$ s eV$^{-1}$ (thin), keeping $\tau_{3}/m_{3} = 10 ^{7}$ s eV$^{-1}$, same as before. The GMF model by Jansson and Farrar \cite{Jansson_2012} is also kept fixed in this case.}
\end{figure*}

In the left panel of Fig. \ref{fig:alplimserr}, we show the upper limits obtained in this case. We find no significant impact on the coupling constant except for a slight improvement in the ALP mass range for both initial neutrino flux models. Thus, the choice of the GMF model does not drastically affect the upper limits. 

\subsubsection{Impact of neutrino decay lifetime}
Next, we discuss the uncertainty due to the assumption of neutrino decay lifetime. We choose $\tau_{2}/m_{2} = 10 ^{5}$ s eV$^{-1}$ keeping $\tau_{3}/m_{3} = 10 ^{7}$ s eV$^{-1}$, same as before. The ALP production probabilities are shown in the left panel of Fig. \ref{fig:alpprodprob} by thin solid and dashed curves. We find that compared to the benchmark value of $\tau_{2}/m_{2} = 10 ^{4}$ s eV$^{-1}$ , the ALP production probability goes down to below $\sim$ 2\% at around 100 GeV. This feature can be clearly seen as a little bump in the corresponding survived neutrino flux shown in the right panel of Fig. \ref{fig:alpprodprob} by the purple dot-dashed curve.

In the right panel of Fig. \ref{fig:alplimserr}, we show the upper limits obtained in this case. We find no significant change in the bound for the benchmark initial neutrino flux, whereas it degrades by a small amount for the initial neutrino flux from model II. However, a slightly bigger region can be excluded in this case. Thus, our limits are not strongly sensitive to the choice of neutrino decay lifetime.

\subsubsection{Impact of $\gamma$-rays to ALP conversion}
Due to nonzero coupling between photons and ALPs, there is a probability that a fraction of the conventional gamma-ray flux will also convert into ALPs in GMF. The converted gamma rays then escape the detection, thus reducing the observed flux. In order to account for this possibility, we multiply the conventional gamma-ray flux with the survival probability $P_{\text{surv}}$ of gamma rays after the ALP conversion,
\begin{equation}
    \Phi^{\text{abs}}(E_{\gamma}) = \Phi^{\text{conv}} (E_{\gamma}) \times P_{\text{surv}}(E_{\gamma})\, .
\end{equation}
We calculate, as an example, the maximum loss of gamma rays due to conversion for the ALP parameters realized in the left panel of Fig. \ref{fig:gamsed} and find that it is around 0.5\% in the relevant energy range from 0.1 GeV to 1 TeV.  We anticipate that this uncertainty will induce a negligible effect on the upper limits. In order to check this, we then again calculate the $\chi^{2}$ using Eq. (\ref{eq:chi2fit}), replacing $\Phi^{\text{conv}}$ with $\Phi^{\text{abs}}$, and $\Delta \chi^{2}$ following Sec. \ref{subsec:alpconstraints}. We find that upper limits remain unaffected by the photon loss due to conversion to ALPs.

\subsection{Diffuse $\gamma$-ray flux at GeV energies under the ALP effect}
In this section, we estimate the contribution to diffuse gamma-ray flux from NGC 1068-like sources at GeV energies by convolving the gamma-ray flux obtained from a single source under the ALP-photon conversion, 
\begin{equation}
    \Phi_{diff}(E_{\gamma}) = \int_{0}^{z_{max}} \frac{d\phi_{a}}{dE} \, n(z) \, \frac{d^{2}V}{dzd\Omega} \, P_{a\gamma}(E_{\gamma}) \, dz \, ,
\end{equation}
where $d\phi_{a}/dE$ is the ALP flux obtained in Eq. (\ref{eq:alpflux}) for NGC 1068, $n(z)$ is the source density that we assume proportional to the star formation rate (SFR) at redshift $z$, $d^{2}V/dzd\Omega$ is the comoving volume element per unit redshift per unit solid angle, and $P_{a\gamma}$ is the conversion probability from ALPs to gamma rays in GMF. The upper limit of integration is taken to be $z_{max}=6$. We consider the functional form of SFR as in Ref. \cite{Yuksel2008}. The parametric fit is taken from Table II of Ref. \cite{horiuchi2009}.

\begin{figure}
    \centering
    \includegraphics[width=\linewidth]{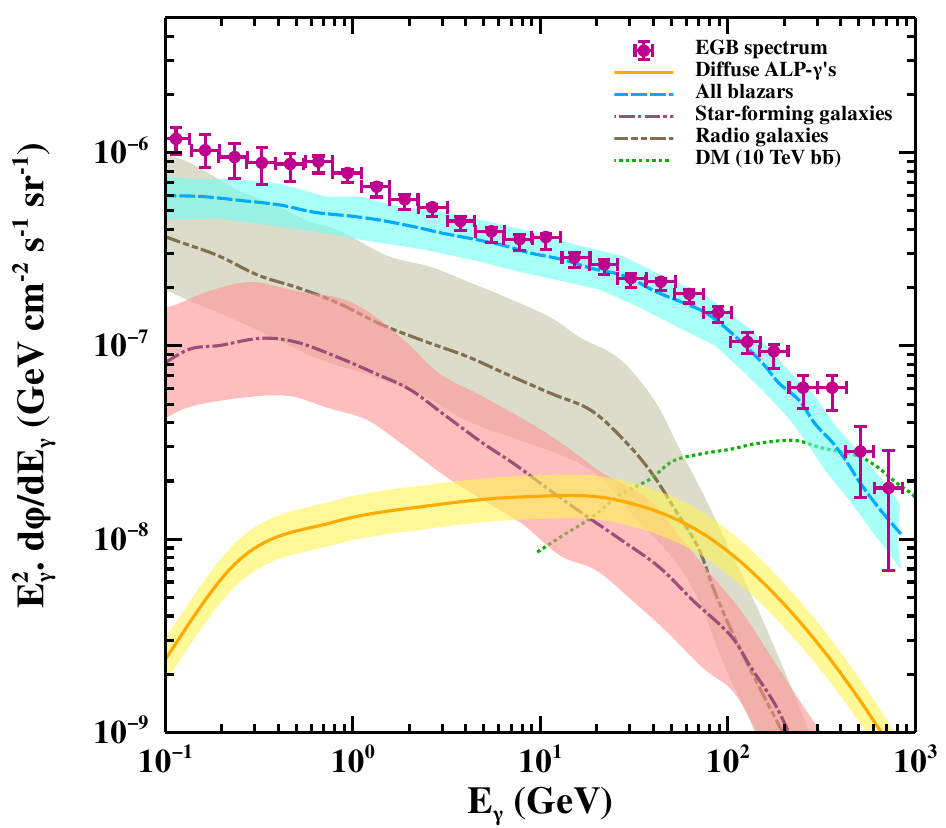}\caption{\label{fig:diffgamma}Expected diffuse gamma-ray flux (orange solid curve)  from invisible neutrino decay to ALPs. The EGB spectrum measured by \textit{Fermi}-LAT \cite{ackermann2015} is shown in purple points. For comparison, we also show the flux from all blazars \cite{ajello2015}, star-forming galaxies \cite{ackermann2012b}, radio galaxies \cite{inoue2011}, and DM-induced gamma rays \cite{ajello2015}.}
\end{figure}

In Fig. \ref{fig:diffgamma}, we show the diffuse gamma-ray flux for benchmark values of neutrino decay lifetime and ALP parameters (m$_a$ = 1 neV, g$_{a\gamma}$ = 1.37 $\times$ 10$^{-11}$ GeV$^{-1}$) along with an 1$\sigma$ uncertainty band. We consider the uncertainty in the parametric fit to the analytical form of SFR (refer to Table II of Ref. \cite{horiuchi2009}) to calculate the lower and upper limits of the 1$\sigma$ uncertainty band. For comparison, we also show the extragalactic gamma-ray background (EGB) measured by \textit{Fermi}-LAT \cite{ackermann2015} and the diffuse flux from all blazars \cite{ajello2015}, star-forming galaxies \cite{ackermann2012b}, radio galaxies \cite{inoue2011}, and dark matter (DM) induced gamma rays as calculated in Ref. \cite{ajello2015}. We find that the cumulative gamma-ray flux is less pronounced below 10 GeV; however, slightly higher flux than star-forming and radio galaxies is obtained above 200 GeV. The overall contribution of gamma-ray flux under the ALP scenario, assuming similar emission from NGC 1068-like sources, is insignificant relative to conventional gamma rays from other populations of extragalactic sources. Thus, the ALP-induced gamma rays are highly suppressed to appear as a distinct signature or account for any significant contribution to EGB flux.

\section{Summary}
The discovery of high-energy astrophysical neutrino sources by IceCube opens up a new opportunity to provide deeper understanding, or possibly discover, and experimentally test new physics beyond the Standard Model. One of the prime quests is to look for signatures of neutrino decay. On the other hand, the induction of spectral irregularities in the high-energy gamma-ray spectra of the astrophysical sources is a promising strategy for searching ALPs. In this work, we have investigated a novel scenario of invisible neutrino decay into ALPs. Our approach offers a complimentary and independent probe to previous studies where gamma rays produced at the source are used to investigate the ALP hypothesis. We assume that a fraction of neutrinos produced at the source undergoes decay while propagating, producing ultralight ALPs before reaching the Earth. These ALPs upon entering into the Milky Way convert into gamma rays due to GMF. These gamma rays will add on top of the conventional gamma rays produced due to hadronic interactions at the source. It might be possible that these ALP-induced gamma rays are sizable enough to leave some imprints in the measured flux. 

We exploited the IceCube and \textit{Fermi}-LAT observations of NGC 1068 to put upper bounds on the ALP parameters. We use the benchmark model for initial neutrino flux at the source given by Ref. \cite{eichmann2022}. Assuming normal mass ordering and the lightest neutrino $\nu_{1}$ to be stable, we consider the benchmark neutrino decay lifetimes of $\tau_2/m_2 = 10^4$ s eV$^{-1}$ and $\tau_3/m_3 = 10^7$ s eV$^{-1}$ satisfying the cosmological constraint by Ref. \cite{barenboim2021}. We obtain an upper limit on the coupling strength at 95 \% CL to be $g_{a\gamma}\lesssim 1.37 \times 10^{-11}$ GeV$^{-1}$ for ALP masses $m_{a} \leq 2 \times 10^{-9}$ eV. Our limits significantly improve over the CAST constraints \cite{cast2017} but are comparable to the constraints set by H.E.S.S. \cite{abramowski2013constraints}, NGC 1275 \cite{ajello2016search}, and Mrk 421 \cite{li2021limits} in this mass range. Using the obtained ALP parameters, we also computed the contribution of ALP-induced gamma rays to the gamma-ray spectra of NGC 1068 by \textit{Fermi}-LAT and MAGIC. Since the conventional gamma rays in the benchmark model fairly explain the observed data points, the contribution of ALP-induced gamma rays is very less to appear as a distinct signature. Thus, only upper bounds could be placed in this case.

We have also discussed the possible systematic uncertainties introduced to the upper limits. First, we study the impact of initial neutrino flux by choosing four different models proposed in the literature. Each model differs in emission region and mechanism to explain the gamma rays and neutrino emission from the source. We found that except for the case of minimal $pp$ interactions in the model given by Ref. \cite{murase2022}, no significant constraint can be put on the coupling strength. For the minimal $pp$ case, we are able to exclude a very narrow region of $9.35 \times 10^{-11}$ GeV$^{-1}$ $ \lesssim g_{a\gamma} \lesssim 1.20 \times 10^{-10}$ GeV$^{-1}$ for ALP masses $m_{a} \le 10^{-9}$ eV. In this model, the conventional gamma rays cannot fully explain the data points above 10 GeV and the ALP-induced gamma rays can fairly explain it. Nonetheless, the obtained ALP parameters have already been ruled out by the CAST experiment. Thus, the choice of initial neutrino flux plays a significant role in constraining the ALP parameters. In the future, a better characterization of the emission mechanism and region using broadband observations will significantly reduce the model uncertainties. Next, we study the impact of choosing a different GMF model by Pshirkov \textit{et al.} over the one given by Jansson and Farrar on the upper limits. We find that the upper limit on the coupling strength is not affected although a slight improvement in the mass range is obtained. We then study the impact of choosing a different neutrino decay lifetimes, $\tau_2/m_2 = 10^5$ s eV$^{-1}$ and keeping $\tau_3/m_3$ same as before, and found that the upper limits are not strongly sensitive to them. Due to the finite probability of absorption of conventional gamma rays due to conversion into ALPs in the GMF, we also study its impact and found that it does not affect the upper limits.

Finally, we estimate the contribution to diffuse gamma-ray flux from NGC 1068-like sources under the ALP scenario. We then compared this flux with the EGB spectrum measured by \textit{Fermi}-LAT in the GeV energies along with the contribution to the diffuse flux from blazars, star-forming galaxies, radio galaxies, and dark matter. We find that ALP-induced gamma rays contribute less significantly compared to other populations of extragalactic sources.

In the near future, the advent of next-generation neutrino telescopes like IceCube-Gen2, KM3NeT, etc., with their unprecedented sensitivities and the rapid increase in multimessenger studies, especially gravitational waves, will allow further exploration of invisible neutrino decay. The implications of which may appear in the gamma-ray band and can be probed by MAGIC, High Energy Stereoscopic System (H.E.S.S.), and Cherenkov Telescope Array (CTA). This will ultimately lead to closing the gaps in the ALP parameter space, which is still unconstrained by the existing bounds.

\begin{acknowledgments}
The author would like to thank the anonymous referee for constructive comments, which helped in improving the quality of the manuscript.
\end{acknowledgments}

% Create the reference section using BibTeX:
\bibliography{references}

\end{document}